\documentclass[fleqn,twoside]{article}
\usepackage{espcrc2}

\usepackage{graphicx}

\newcommand{\beq}{\begin{equation}}
\newcommand{\eeq}{\end{equation}}
\newcommand{\beqn}{\begin{eqnarray}}
\newcommand{\eeqn}{\end{eqnarray}}

\title{
\vspace{-9mm} \rightline{\small ITEP-LAT/2003-20} \vspace{-2mm}
\rightline{\small September, 2003}
Geometry of the monopole clusters at different scales
    \thanks{Talk presented by P.Yu.B. at Lattice 2003, Tsukuba}
  }

\author{
    V.G. Bornyakov\address[IHEP]{Institute for High Energy Physics,
        Protvino 142284, Russia},
        P.Yu. Boyko\address[ITEP]{Institute of Theoretical and  Experimental
    Physics, B.~Cheremushkinskaya~25, Moscow, 117259, Russia},
        M.I. Polikarpov\addressmark[ITEP]
        and
V.I. Zakharov \address[MPI]{Max-Planck Institut f\"ur Physik, F\"ohringer Ring
6, 80805, M\"unchen, Germany}\thanks{Work is partially supported by grants RFBR
02-02-17308, \uppercase{RFBR 01-02-17456, DFG-RFBR 436 RUS 113/739/0,
INTAS-00-00111} and \uppercase{CRDF} award \uppercase{RPI-2364-MO}-02.}
}

\begin{document}

\begin{abstract}
We present results of measurements of various geometrical characteristics
of the monopole clusters in the maximally Abelian projection of SU(2) lattice
gauge theory. We observe scaling for the observables tested. 
Short clusters correspond to random
walks at small scale but have long-range correlations at the hadronic scale. 
On the other hand, the percolating cluster at the hadronic scale 
does not correspond to a random walk. 

\vspace{1pc}
\end{abstract}

\maketitle

%-----------------------------------------------------------------------------
\section{INTRODUCTION}

The monopoles of the maximal Abelian projection of
SU(2) gluodynamics seem to be  adequate dynamical variables 
for the description of
confinement mechanism (for a review and references see, e.g.\cite{review})
and further investigation of their properties is
important. The monopoles are observed as trajectories 
closed on the dual lattice. Moreover, these trajectories form two
types of clusters, the percolating cluster and finite ones,
for a review and references see, e.g. ~\cite{HT}.  In particular,
the percolating
cluster, spreading over the whole of the lattice is
responsible for the confinement. 

Although phenomenology of the monopoles is quite rich,
there is no understanding of the nature of the monopoles
on the fundamental level. Thus, further accumulation of
the data on the monopoles and tests of various models seems to
be the best strategy available now. Here we report on observation of 
scaling properties of various geometrical characteristics of the
monopole clusters. We perform calculations on various lattices at
different values of $\beta$, the number of gauge fields configurations
is given below in Table~2. Further details can be found in
~\cite{lat02,mcl-nph}.

In Sect.~\ref{s2} we discuss the scaling
behavior of the various characteristics of the percolating monopole cluster.
The geometrical properties of finite clusters are presented in Sect.~\ref{s3}.
The density of clusters is discussed in Sect~\ref{s4}. Conclusions are given in
Sect.~\ref{s5}. 
%-----------------------------------------------------------------------------
\section{PERCOLATING CLUSTER GEOMETRY}\label{s2}
\subsection{Segments and Crossings} The percolating cluster consists of
segments (that is, trajectories between crossings) and crossings.  
The natural parameters to be measured
are the average length of the segment, $\langle l_{\rm segm} \rangle$  and the
average Euclidean distance between the end points of the segments,
$\langle d \rangle$.
The data on $\langle l_{\rm segm} \rangle$ we obtained demonstrate
scaling \cite{lat02,mcl-nph}:
\beq \langle l_{\rm segm}\rangle ~\approx~ 1.60~fm~~. \eeq
The value of $\langle d \rangle$ exhibits stronger
variation with the lattice spacing $a$ and can be fitted as:
\beq \langle d \rangle ~\approx~ \big(0.3 ~+~(a/fm~-~0.11)\big)~fm~,
\eeq
for the values of $a$ tested, $0.06~fm<~a~<~0.16~fm$.
%These values can be compared with two typical scales of the monopole currents:
%average radius $\langle\rho_m\rangle \approx 0.05~fm$ and the average
%inter-monopole distance $\langle R\rangle \approx 0.5~fm$ \cite{anatomy}.

\begin{figure}
\begin{center}
\includegraphics[width=\columnwidth]{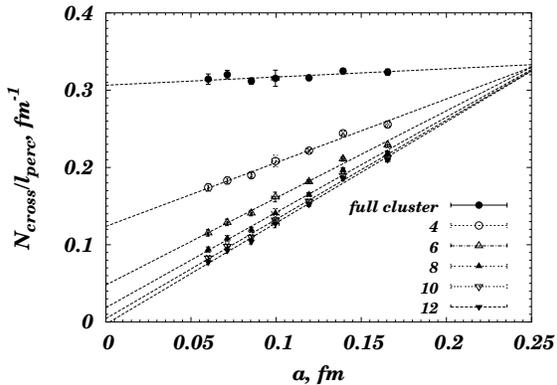}
\vspace{-12mm}
\caption {Number of crossings per unit physical length vs. lattice spacing in
full percolating cluster and in the percolating cluster with excluded closed
loops. Numbers $4, 6, ... 12$ in the legend mean that closed loops of the length up to $4, 6, ... 12$ are
excluded. The dashed lines show the linear fits.
\vspace{-7mm} 
} 
\label{Ncrossings_fig}
\end{center}
\end{figure}

Scaling of the average segment length implies scaling of the number of the
cluster self-crossings per unit physical length of the monopole trajectory,
$N_{cross}/l_{perc}$. In Fig.~\ref{Ncrossings_fig} the number of crossing 
points per unit
length of the monopole trajectory is shown by filled
circles. According
to direct measurements \cite{lat02,mcl-nph}:
$N_{cross}/l_{perc}\approx 0.3\, fm^{-1}$.
This number  changes if we exclude the closed loops of finite length (in
lattice units) connected to the percolating cluster. The number of crossings
reduces and is compatible with zero in the continuum limit if we exclude the
loops of the length up to 8 (or up to larger length), see
Fig.~\ref{Ncrossings_fig}.

\subsection{Angular Correlations} On a hypercubic lattice one can
measure three different angular correlations between
tangents to the monopole trajectory. Namely, 
two links connected by $l$ cluster links
may have the same, the opposite and neither the same nor the opposite ("the
other") directions. We normalize these probabilities to unity for the random
walk case. It was reported in \cite{lat02,mcl-nph} that the probability 
$P_{\rm other}$ is 
equal to unity within error bars for $l$ larger than several lattice spacings 
while the probability $P_{\rm same}$ converges exponentially to unity:
\beq P_{\rm same}~-1~=~ A_s e^{-\mu_s la}\, . \eeq
By definition the sum of all three probabilities is equal to 3. 
This means that 
the difference $P_{\rm opposit}~-1~$  falls off
exponentially with the same (up to statistical errors) decay parameter
and amplitude:
\beq P_{\rm oppos}~-1~=~- A_o e^{-\mu_o la}\, . \eeq
\beq \mu_s~\approx~\mu_o\,,\,\,\, A_s~\approx~A_o\,. \eeq

As it is seen from the Table~\ref{mu} this is indeed true for $\mu_s$ and 
$\mu_o$, which are also independent of
$\beta$ within the errors, and thus correspond to the continuum limit.
\vspace{-5mm}
\begin{table}[!h]
\begin{center}
\caption{The fitted masses in $Mev$ from data for $(P_{\rm same} - 1)$ and $(1
- P_{\rm opposit})$
}
\begin{tabular}{c l l l l}
\hline
$\beta$ & 2.45 & 2.50 & 2.55 & 2.60 \\
$ \mu_{s}$ & 290(60) & 290(20) & 271(15) & 273(12) \\
$\mu_{o}$ & 250(60) & 240(20) & 252(15) & 277(15) \\
\hline
\vspace{-20mm}
\end{tabular}
\label{mu}
\end{center}
\end{table}

\section{FINITE CLUSTERS GEOMETRY}\label{s3}
A set of finite clusters is characterized by its spectrum, an average number of
clusters as the function of their length, $l$. First measurements of the
spectrum were reported for one value of $\beta$ in ref. \cite{HT} and it was
found that the spectrum has $1/l^3$ form. Our data confirm this behavior for
all the values of $a$ considered, see
Table~\ref{tbl2}. As for the cluster radius, we observe:
\beq \langle r \rangle ~\sim~ \sqrt{l\cdot a}\, .\eeq
For interpretation of the data on finite clusters see
\cite{chernodub,mcl-nph}.
\begin{table*}
\caption{The fit of the length spectrum of the finite clusters by the function
$1/l^\alpha$ , for various values of $\beta$ on the lattices $L^4$; $N_{conf}$
is the number of considered gauge fields configurations at given $\beta$ and
$L$.}
\begin{tabular}{ c l l l l l l l l l l l}
\hline
$\beta$    & 2.30 & 2.35 & 2.40 & 2.40 & 2.40 & 2.45 & 2.50  & 2.55 & 2.60 \\
L          &  16  & 16   & 16   & 24   & 32   &  24  & 24    & 28   & 28 \\
$N_{conf}$ & 100  & 100  & 300  & 137  & 35   &  20  & 50    & 40   & 50 \\
$\alpha$ & 3.12(4) & 3.10(4) & 2.98(2) & 2.95(2)  & 2.97(2) & 2.91(3) & 3.02(3) &
3.06(3) & 3.11(4)\\
\hline
\vspace{-8mm}
\end{tabular}
\label{tbl2}
\end{table*}

\section{DENSITY OF CLUSTERS} \label{s4} The standard definition of the
densities of the percolating and the finite clusters is:
\beqn
\label{l} <l_{\rm perc}>~\equiv~4~\rho_{\rm perc} a^4 N_{\rm sites}\,, \\
<l_{\rm fin}>~\equiv~4~\rho_{\rm fin} a^4 N_{\rm sites}\,,
\eeqn
where $<l_{\rm perc}>$ and $<l_{\rm fin}>$ is the average length of the
corresponding clusters, $N_{\rm sites}$ is the number of the lattice sites.

\begin{figure}
\begin{center}
\includegraphics[width=\columnwidth]{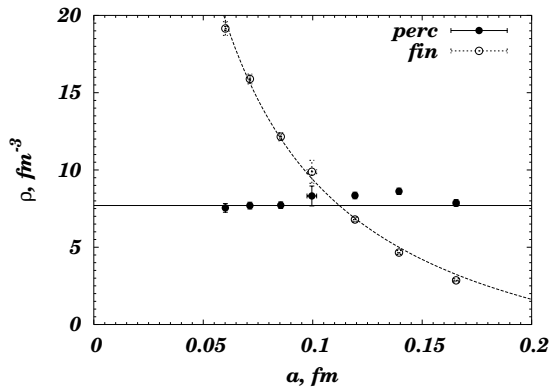}
\vspace{-12mm}
\caption{Density of the finite clusters $\rho_{fin}$ and percolating clusters
$\rho_{\rm perc}$; solid and dashed lines, are fits by a constant  and
 function (\ref{rhofin})  respectively.
\vspace{-8mm}
} \label{rho_fig}
\end{center}
\end{figure}

In Fig.~\ref{rho_fig} we show the monopole density $\rho_{\rm perc}$ as a
function of the lattice spacing $a$. Our results are in agreement with those of
ref.~\cite{bmp} but obtained with higher statistics. The dependence on $a$ is
rather weak and the fit of the data by a constant for $\beta > 2.35$ gives:
\begin{equation}\label{rhoperc}
\rho_{{\rm perc}}~=~7.70(8)\, fm^{-3}~~.
\end{equation}

The density of the finite clusters can be
perfectly fitted (dashed curve on Fig.\ref{rho_fig}) as
\begin{equation}\label{rhofin}
\rho_{\rm fin}~ =~ C_1~ + ~\frac{C_2}{a}\, ,
\label{fit_df}
\end{equation}
where $C_1 = -6.1(5)\,fm^{-3}~,~ C_2~ =~ 1.55(4) \, fm^{-2}$. The negative
value of the constant $C_1$ means that the fit is not valid for large 
(unphysical) values of
the lattice spacing.
%-----------------------------------------------------------------------------
\section{CONCLUSIONS}
\label{s5}

The observed scaling behavior of geometrical characteristics of the
monopole clusters supports a conjecture that the monopoles 
might correspond to gauge invariant
objects. Most remarkable, the scaling behaviour holds for small $a$.
With implication that the monopole size can be very small, see also
\cite{anatomy}. Moreover, the $1/a$ divergence of the finite clusters density 
means that these clusters are associated with two dimensional surfaces
embedded into the four dimensional space~\cite{mcl-nph}. This relation between monopoles and surfaces is in a sense very close to the relation between P-vortices and monopoles observed in ref.~\cite{zelcite}, see also discussion in the talk presented by
S.N.~Syritsyn at this conference~\cite{syritsyn}.


\begin{thebibliography}{99}
\bibitem{review}
M.N.~Chernodub, M.I.~Polikarpov,~in~"Confinement,~duality,~and nonperturbative
aspects of QCD", p.~387, ed. Pierre van Baal, Plenum Press, 1998,
hep-th/9710205.

\bibitem{HT} A.~Hart and M.~Teper, Phys. Rev. D58 (1998) 014504,
Phys. Rev. D60 (1999) 114506, {\tt hep-lat/9902031}.

\bibitem{lat02}
P.Yu.~Boyko, M.I.~Polikarpov and V.I.~Zakharov, Nucl. Phys. (Proc. Suppl.) 
119 (2003) 724 {\tt hep-lat/0209075}.

\bibitem{mcl-nph}
V.G. Bornyakov P.Yu.~Boyko, M.I.~Polikarpov and V.I.~Zakharov, ''Monopole
clusters at short and large distances'' {\tt hep-lat/0305021}.

\bibitem{chernodub}
M.N. Chernodub and V.I. Zakharov, ''Towards understanding structure
of the monopole clusters'', {\tt hep-th/0211267}.

\bibitem{bmp}
V.~Bornyakov and M.~M\"uller-Preussker,
 Nucl.\ Phys.\ (Proc.\ Suppl.\ )  106 (2002) 646,  {\tt hep-lat/0110209}.

\bibitem{anatomy}
V.G.~Bornyakov {\rm et. al.},  Phys. Lett.  B537 (2002) 291, {\tt
hep-lat/0103032}.


\bibitem{zelcite}
J. Ambjorn, J. Giedt and J. Greensite,  JHEP  9903 (1999) 019.


\bibitem{syritsyn}
A. V. Kovalenko, M. I. Polikarpov, S. N. Syritsyn and V. I. Zakharov,
``Interplay of Monopoles and P-Vortices'', proceedings of this conference.


\end{thebibliography}
\end{document}